\RequirePackage[2022-02-02]{latexrelease}
\documentclass[%
 reprint,
superscriptaddress,
 amsmath,amssymb,
 aps,
]{revtex4-2}

\newcommand{\beorn}{{\tt BEoRN}}
\newcommand{\dTb}{$dT_{b}$}
\newcommand{\dTbbar}{$\bar{dT}_{b}$}
\newcommand{\xHII}{$x_{\rm HII}$}
\newcommand{\xHIIbar}{$\bar{x}_{\rm HII}$}
\newcommand{\xHI}{$x_{\rm HI}$}
\newcommand{\xal}{$x_{\rm \alpha}$}
\newcommand{\xalbar}{$\bar{x}_{\rm \alpha}$}
\newcommand{\db}{$\delta_{\rm b}$}
\newcommand{\da}{$\delta_{\rm \alpha}$}
\newcommand{\dT}{$\delta_{\rm T}$}
\newcommand{\dr}{$\delta_{\rm r}$}
\newcommand{\lyal}{Lyman-${\rm \alpha}$}
\newcommand{\Tk}{$T_{\rm k}$}
\newcommand{\Tkbar}{$\bar{T}_{\rm k}$}

\newcommand{\xcl}{$x_{\rm cl}$}

\newcommand{\cutoff}{\textit{cutoff}}
\newcommand{\default}{\textit{default}}
\newcommand{\boost}{\textit{boost}}
\newcommand{\U}{$U_{\rm \alpha}$}
\newcommand{\V}{$V_{\rm k}$}

\newcommand{\Utayl}{$U_{\rm \alpha,\, taylor}$}
\newcommand{\Vtayl}{$V_{\rm k,\, taylor}$}
\newcommand{\dastar}{$\delta^{*}_{\rm \alpha}$}

\usepackage{graphicx}
\usepackage{dcolumn}
\usepackage{bm}
\usepackage{hyperref}

\usepackage{xcolor}

\begin{document}

\preprint{APS/123-QED}

\title{Testing common approximations to predict the 21cm signal at the Epoch of Reionization and Cosmic dawn}

\author{Timoth\'ee Schaeffer}
\affiliation{Department of Astrophysics, University of Zurich, Winterthurerstrasse 190, 8057 Zurich, Switzerland.}
\email{timothee.schaeffer@uzh.ch}

\author{Sambit K. Giri}
\affiliation{Nordita, KTH Royal Institute of Technology and Stockholm University, Hannes Alf\'vens v\"ag 12, SE-106 91 Stockholm, Sweden}%

\author{Aurel Schneider}%
\affiliation{Department of Astrophysics, University of Zurich, Winterthurerstrasse 190, 8057 Zurich, Switzerland.}
\email{aurel.schneider@uzh.ch}

\date{\today, NORDITA 2024-004 
}

\begin{abstract}
Predicting the 21cm signal from the epoch of reionization and cosmic dawn is a complex and challenging task. Various simplifying assumptions have been applied over the last decades to make the modeling more affordable. In this paper, we investigate the validity of several such assumptions, using a simulation suite consisting of three different astrophysical source models that agree with the current constraints on the reionization history and the UV luminosity function. We first show that the common assumption of a saturated spin temperature may lead to significant errors in the 21cm clustering signal over the full reionization period. The same is true for the assumption of a neutral universe during the cosmic dawn which may lead to significant deviation from the correct signal during the heating and the Lyman-$\alpha$ coupling period. Another popular simplifying assumption consists of predicting the global differential brightness temperature ($dT_b$) based on the average quantities of the reionization fraction, gas temperature, and Lyman-$\alpha$ coupling. We show that such an approach leads to a 10 percent deeper absorption signal compared to the results obtained by averaging the final $dT_b$-map. Finally, we investigate the simplifying method of breaking the 21cm clustering signal into different auto and cross components that are then solved assuming linearity. We show that even though the individual fields have a variance well below unity, they often cannot be treated perturbatively as the perturbations are strongly non-Gaussian. As a consequence, predictions based on the perturbative solution of individual auto and cross power spectra may lead to strongly biased results, even if higher-order terms are taken into account.

\end{abstract}

\maketitle


\section{Introduction}
The cosmic dawn and epoch of reionization (EoR) designate the periods from the emergence of the first stars and galaxies to the completion of the reionization process. During this phase, the light from these sources gradually penetrated the inter-galactic medium (IGM), modifying the spin distributions, heating, and eventually ionizing the neutral hydrogen (HI) atoms. The resulting fluctuations in the temperature and ionization fraction of the IGM leave distinctive features on the hyperfine 21-cm signal emitted by neutral hydrogen. Consequently, the 21cm signal serves as a powerful probe, sensitive to the properties of the first sources of light \cite{Mirocha:2017xxz, Mebane:2020jwl, Magg:2021jyc, Kulkarni:2017qwu, Ventura:2022rwn, Fialkov:2019vnb, Reis:2020arr}, to the cosmological parameters \cite{McQuinn:2005hk, Mao:2008ug, Liu:2015gaa, Kern:2017ccn, HMreio}, and potential extensions to the standard $\Lambda$-cold dark matter ($\Lambda$CDM) model \cite{Lopez-Honorez:2018ipk, Driskell:2022pax, Barkana:2022hko, Lopez-Honorez:2017csg, Schneider:2018xba, Munoz:2020mue, Giri:2022nxq}.

The 21cm signal is targeted by various ongoing or planned surveys. While it has not been detected yet, experiments such as the Low-Frequency Array \citep[LOFAR,][]{vanHaarlem:2013dsa}, the Murchison Widefield Array \citep[MWA,][]{Tingay:2012ps}, the Hydrogen Epoch of Reionization Array \cite[HERA,][]{DeBoer:2016tnn}, \citep[GMRT,][]{GMRT2013:uplim}, and the Precision Array for Probing the Epoch of Reionization \citep[PAPER,][]{PAPER_2019_uplim} have provided upper limits for the 21cm power spectrum. 
These upper limits have already been used to rule out some regions of parameter space populated with rather extreme models \citep{Ghara:2020syx,ghara2021constraining,MWA_constr_Greig,greig2021interpreting,HERA:2021noe} placing lower bounds on the normalization of the X-ray spectrum of high redshift sources. Moreover, forecast studies have shown the potential of the 21cm power spectrum to constrain cosmological parameters with precision competitive with other probes such as the cosmic microwave background (CMB) radiation \cite{McQuinn:2005hk, Mao:2008ug,Liu:2015gaa,Kern:2017ccn,HMreio}. These results affirm the significant potential of the 21cm signal to provide complementary constraints on cosmological models from an entirely new redshift window.

The task of extracting physical information from the 21cm signal presents considerable challenges. Not only is the unknown astrophysical and cosmological parameter space vast, but the signal is also difficult and computationally expensive to simulate accurately. It requires modeling the
formation of galaxies down to the smallest star-forming
halos, resolving the processes through which light escapes
the interstellar medium and reaches the IGM, propagating this light across large cosmological distances, while
simultaneously solving coupled radiative transfer equations to track its interaction with the IGM gas. See Ref.~\citep{iliev2006cosmological,iliev2009cosmological,Grizzly_comparison_3D_1D,gnedin2022modeling} for a more detailed discussion about these processes. 

Given the high computational costs of radiative-transfer simulations, statistical inference of data is often performed with fast semi-numerical or analytical methods \citep[e.g.,][]{Greig:2015qca,HaloModel_Paper,Giri:2022nxq,HMreio,monsalve2019results}. These approaches rely on assumptions and approximations which may lead to errors in the predicted signal.
The uncertainty of a method can be quantified with a theory or modeling error (which may be a redshift and scale-dependent quantity). See Ref.~\citep{Greig:2015qca} for a study of this error. It is usually introduced in statistical inference pipelines as an additional error, added in quadrature to the covariance matrix, weakening the constraining power of the analyzes. As shown in Ref.~\citep{HMreio}, it is crucial to reduce this error to be able to produce competitive parameters inference with future 21cm data. 


In this paper, we test different assumptions commonly made to predict the 21cm global signal and power spectrum. Using the one-dimensional radiative transfer code \beorn{} \citep{Schaeffer:2023rsy}, we generate a set of simulations with three different astrophysical source models that all agree with current observations of the reionization fraction and the UV luminosity function. These simulations provide 3-dimensional grids of the density field, the ionization fraction, the kinetic temperature of the gas, the \lyal{} flux, and the 21cm brightness temperature, between redshift $z=25$ and $z=6$. We utilize these simulations to check the validity of various approximations regularly done in the literature.

The paper is structured as follows. In Section~\ref{sec:21cm}, we review the fundamental equations governing the 21cm signal and introduce our suite of simulations. In Sec.~\ref{sec:3}, we investigate the impact of neglecting reionization during cosmic dawn on the signal. Additionally, we test the validity of the saturated spin temperature assumption during reionization. In Sec.~\ref{sec:testing PT}, we describe the perturbative approach to compute the 21 cm power spectrum and investigate its validity. Finally, Sec.~\ref{sec:conclusions} summarizes our findings. 

Throughout this paper, bar symbols above a letter denote spatial average. For any field $X$, we define the normalized fluctuation $\delta_{\rm X}=X/\bar{X}-1$. 
We will assume cosmological parameters consistent with Planck 2018 results \citep{Planck:2018vyg}, setting the matter abundance $\Omega_{\rm m}=0.31$, baryon abundance $\Omega_{\rm b}=0.045$, and dimensionless Hubble constant $h=0.68$. The standard deviation of matter perturbations at 8$h^{-1}$ cMpc scale is $\sigma_{\rm 8}=0.81$.

\section{21cm signal : theory and modeling}\label{sec:21cm}


The 21cm signal emitted by neutral hydrogen (HI) during the cosmic dawn and the EoR promises to be a powerful probe of cosmology and astrophysics. This signal is targeted by radio interferometers such as the SKA, which are sensitive to the differential brightness temperature \dTb{}. The evolution of \dTb{} follows the relation \citep{Furlanetto:2006tf}
\begin{align}\label{eq:dTb}
    dT_{b}(\mathbf{x},z)&\simeq T_0(z) x_{\rm HI}(\mathbf{x},z)\left[1+\delta_b(\mathbf{x},z)\right]\\
    &\times U_{\rm \alpha}(\mathbf{x},z) V_{\rm k}(\mathbf{x},z),  \nonumber
\end{align}
with the amplitude of the signal $T_{0}$ given by
\begin{eqnarray}
T_0(z)=27 \left(\frac{\Omega_bh^2}{0.023}\right)\left(\frac{0.15}{\Omega_mh^2}\frac{1+z}{10}\right)^{\frac{1}{2}}\,\,\, {\rm mK},
\end{eqnarray}
where $\Omega_m$ and $\Omega_b$ are the cosmic matter and baryon abundances and $h=H_0/100$ (km/s)/Mpc is the dimensionless Hubble parameter.

\textcolor{black}{The quantities $U_{\rm \alpha}$ and $V_{\rm k}$ are defined by
\begin{equation}\label{eq: UV def}
    U_{\rm \alpha}\times V_{\rm k} = \left[1-\frac{T_{\rm cmb}(z)}{T_{\rm S}(\mathbf{x},z)}\right]
\end{equation}
with $T_{\rm S}$ the spin temperature of neutral hydrogen given by
\begin{equation}\label{eq: Tspin}
    T_{\rm S}^{-1}(\mathbf{x},z)= \frac{T^{-1}_{\rm cmb}(z)+x_{\rm tot}(\mathbf{x},z)T^{-1}_{\rm k}(\mathbf{x},z)}{1+x_{\rm tot}(\mathbf{x},z)}
\end{equation}
where $x_{\rm tot} = x_{\alpha}(\mathbf{x},z)+x_{\rm cl}(\mathbf{x},z)$. Plugging Eq.}\ref{eq: Tspin} in Eq.\ref{eq: UV def}, we obtain $U_{\rm \alpha}$ and $V_{\rm k}$ as separate non-linear functions of the \lyal{} coupling coefficient \xal{} and the kinetic temperature \Tk{}, respectively:
\begin{equation}\label{U}
    U_{\rm \alpha} = \frac{x_{\rm tot}(\mathbf{x},z)}{1+x_{\rm tot}(\mathbf{x},z)},
\end{equation}
and
\begin{equation}\label{V}
    V_{\rm k} = \left[1-\frac{T_{\rm cmb}(z)}{T_{\rm k}(\mathbf{x},z)}\right]
\end{equation}
The neutral fraction (\xHI), the baryon overdensity (\db), the \lyal{} coupling coefficient (\xal), the collisional coupling coefficient (\xcl), and the gas temperature (\Tk) are all position ($\mathbf{x}$) and redshift-dependent ($z$). We assume the radio background to be dominated by the homogeneous CMB with temperature $T_{\rm cmb}(z)$. Several studies have explored the possibility of an excess radio background beyond the CMB \citep[e.g.,][]{mondal2020,ghara2021constraining,he2023inverse}, which could also be a position-dependent quantity \citep{Reis:2020arr}. However, we will not consider such a signal in this study. The coefficients \xal{} and \xcl{} are given by:
\begin{equation}
  x_{\rm \alpha}(\mathbf{x},z) = \frac{1.81\times 10^{11}}{(1+z)}S_{\alpha}J_{\alpha}(\mathbf{x},z)
  \label{xal},
\end{equation}
and
\begin{equation}
  x_{\rm c}(z) = \frac{T_{*}}{A_{10} T_{\gamma}(z)}\underset{i=H,e^{-}}{\sum} n_{i}(z)\kappa_{10}^{i}(T_{\rm k}),
  \label{xcol}
\end{equation}
where $J_{\alpha}(\mathbf{x},z)$ is the local flux of Lyman$-\alpha$ photons, $S_{\alpha}$ is given by Eq.~(55) in \citet{Furlanetto:2006jb}.
$\kappa^{i}_{10}$ is the rate coefficient for spin de-excitation in collisions with species $i$ with density $n_{i}$. $A_{10}[\rm s^{-1}]$ is the Einstein coefficient for spontaneous emission, and $T_{*} = 68$ mK the temperature of the hyperfine transition. 

According to Eq.~(\ref{eq:dTb}), \dTb{} is a multi-linear function of \xHI,  \db, $U_{\rm \alpha}$ and $V_{\rm k}$, and a non-linear function of \Tk{} and \xal. It fluctuates between regions ionized by UV photons where $dT_{\rm b}=0$, cold adiabatically cooling regions where the signal is seen in absorption ($dT_{\rm b}<0$), and regions heated by X-ray photons above the CMB temperature where the signal is seen in emission ($dT_{\rm b}>0$). The temporal evolution of the morphology of these regions contains valuable information about the distribution and properties of the first stars and galaxies responsible for heating and ionizing the IGM.

Two summary statistics are commonly used to compress the information contained in the sky data. The first one is the global 21cm signal, defined as the mean value of the \dTb{} field, computed over a sample volume V:
\begin{equation}
  \bar{dT_{\rm b}}(z) = \frac{1}{V} \int_{V} dT_{b}(\mathbf{x},z) d\mathbf{x}
  \label{eq:mean dTb definition}
\end{equation}
the second is the spherically averaged power spectrum $P_{21}(k,z)$ of the \dTb{} field.

We define the power spectrum $P_{\rm F}(k)$ of a given field $F(x)$ as
\begin{multline}\label{eq:PS def}
\big\langle F(k) F^{*}(k')\big \rangle = (2\pi)^{3}\delta^{3D}(k-k') P_{F}(k),
\end{multline}
which means that the total 21cm power spectrum becomes
\begin{multline}\label{eq:21cm PS def}
\big\langle dT_{\rm b}(k) dT^{*}_{\rm b}(k')\big \rangle = (2\pi)^{3}\delta^{3D}(k-k') P_{21}(k),
\end{multline}
where $\delta^{3D}$ is the three-dimensional Dirac delta. Note that the definition of the 21cm power spectrum may vary among different studies. In \citep{Schaeffer:2023rsy}, we introduced $P_{\rm tot}$, defined as the power spectrum of the normalized fluctuation $\delta_{\rm dT_{\rm b}}(x)$. Subsequently, we plotted the quantity $\bar{dT_{\rm b}}^{2}P_{\rm tot}$, which is equivalent to $P_{21}$ as defined above. In the present paper, we use the definition of $P_{21}$ instead, which corresponds to the quantity measured by radio interferometers. Given a power spectrum $P(k,z)$, we introduce its counterpart $\Delta^{2}(k,z) = k^{3}P(k,z)/(2\pi^{2})$, which is independent of length dimension. 


Various techniques exist to model the 21cm signal. They can be broadly put into two categories: (i) grid-based methods and (ii) analytical methods not based on a grid. In the following sections, we will describe a subset of both of these approaches.

\subsection{Simulations over cosmological volumes}\label{sec:simulations}
Numerous grid-based approaches have been developed with the primary objective of simulating the 21cm signal. They include the excursion-set-based codes (e.g., \texttt{21cmFAST} \citep{Mesinger:2011aaa}, \texttt{SimFast21} \citep{santos2010fast}, \texttt{CIFOG} \citep{Hutter_2018}, see also \citep{fialkov2014rich}), hydrodynamic-radiative-transfer frameworks such as \texttt{Licorice} \citep{Semelin_2007,meriot2024loreli}, radiative transfer codes designed to post-process N-body simulations (e.g., the numerical scheme from Ref.~\citep{SOKASIAN2001} or Ref.~\citep{Mittal_23}, \texttt{CRASH} \citep{maselli2003crash}, and \texttt{pyC$^2$RAY} \citep{MELLEMA_2006,hirling2023pyc}), as well as 1-dimensional radiative transfer methods (e.g., \texttt{Bears} \citep{thomas2009fast}, \texttt{Grizzly} \citep{Grizzly_comparison_3D_1D} and \beorn{} \citep{Schaeffer:2023rsy}). These methods are all designed to compute the evolution of \Tk{}, \xHI{}, \xal{}, and \dTb{} on a discretized grid. Then, the mean and the power spectrum of \dTb{} are computed directly from the map using Eqs.~\ref{eq:mean dTb definition} and \ref{eq:21cm PS def}, respectively.

The present analysis relies on grid-based simulations performed with the code \beorn{}, which was introduced and validated in \citep{Schaeffer:2023rsy}. We provide an overview of the main ingredients and methodology of the code in Sec.~\ref{sec:beorn}. Our results will be systematically presented for three different astrophysical source models detailed in Sec.~\ref{sec:bench models}.

\subsubsection{BEoRN}\label{sec:beorn}
\beorn{} is a publicly available Python code \citep{Schaeffer:2023rsy} designed to generate cosmological boxes of 21cm differential brightness temperature \dTb{} throughout the cosmic dawn and EoR \footnote{\beorn{} is publicly available on: \url{https://github.com/cosmic-reionization/BEoRN}.}. It is based on a simple one-dimensional radiation profile approach developed in \citep{Schneider:2020xmf}. \beorn{} reads in halo catalogs and density fields from a pre-run $N$-body simulation to construct the $dT_{\rm b}$ signal on a grid. It populates halos with galaxies according to a flexible source model. The mass accretion rate of halos $\dot{M}_{\rm h}$ is related to the galaxy star formation rate $\dot{M}_{\rm *}$ via a parameterized stellar-to-halo function $f_{\rm *}=\dot{M}_{\rm *}/\dot{M}_{\rm h}$. Additionally, the spectral energy distribution of galaxies is parameterized independently in the X-ray, \lyal{}, and ionizing photon energy bands. 

For a given set of source model parameters, \beorn{} solves 1-dimensional radiative transfer equations to compute profiles for the temperature, the \lyal{} flux, and the size of ionized bubbles around galactic sources. These profiles are then painted onto a grid around halo centers, and the overlap of ionized bubbles is managed consistently by redistributing the excess photons around the boundaries of the connected ionized regions. In that manner, \beorn{} produces 3-dimensional maps of the ionized hydrogen fraction \xHII, the \lyal{} coupling coefficient \xal, the kinetic temperature \Tk, and the brightness temperature \dTb{} over cosmological volumes at various redshifts. We refer to \citep{Schaeffer:2023rsy} for more details regarding the source model parameters and the equations underlying the profiles.

\begin{figure*}
    \centering
    \includegraphics[width =1\textwidth,trim=0.cm 0.cm 0cm 0.0cm,clip]{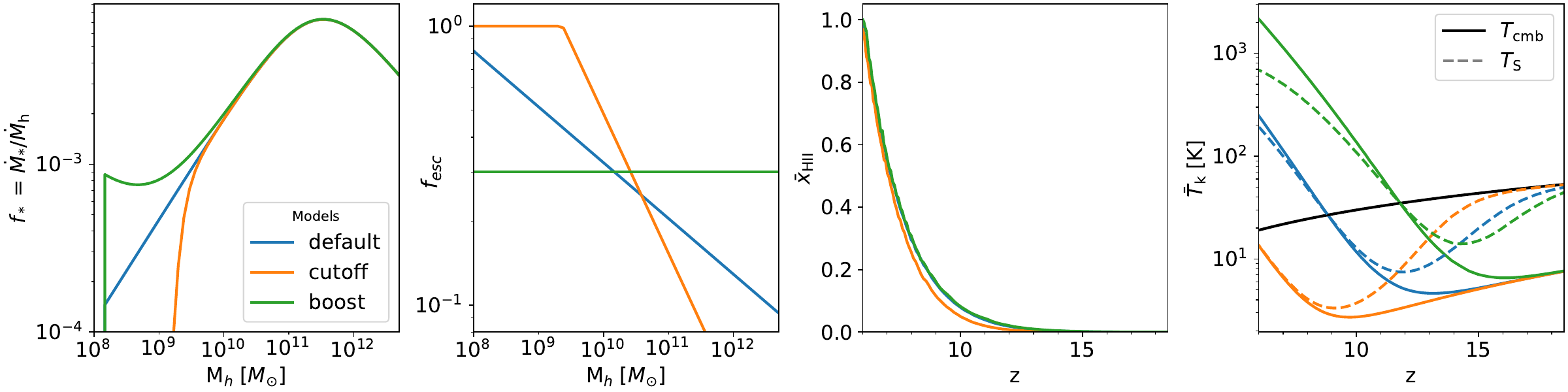}
    \caption{\textbf{Summary of the characteristics of the three benchmark models.} The  \default, \cutoff, and \boost{} models are represented in blue, orange, and green, respectively. \textit{Leftmost panel}: Stellar-to-halo relation ($f_*$) as a function of halo mass. Combined with our halo catalogs, each $f_*$ leads to UV luminosity functions in agreement with observations. \textit{Second panel}: Escape fractions ($f_{\rm esc}$) of ionizing photons as a function of halo mass. We have tuned $f_{\rm esc}$ in each model to obtain reionization histories consistent with observations.  \textit{Third panel}: Mean ionization fraction history. \textit{Rightmost panel}: \textcolor{black}{The mean kinetic temperature and the average spin temperature of the neutral gas are shown as solid and dashed coloured lines, respectively}. The black solid line corresponds to the CMB temperature. The different heating histories arise from the different $f_{*}$ as well as the varying normalization of the X-ray amplitude in each model.
    }
    \label{fig:properties 3models}
\end{figure*}

\subsubsection{The three benchmark models}\label{sec:bench models}
The precise properties of high-redshift galaxies, including their abundance and spectral properties, remain largely unknown, leaving some freedom in the choice of astrophysical parameters. To explore the dependency of our conclusions on astrophysical assumptions, we will perform our analysis for three different benchmark source models, called \cutoff, \default, and \boost, which were introduced in \citep{Schneider:2020xmf,Schaeffer:2023rsy}. These three models are characterized by different stellar-to-halo relation $f_{\rm *}$, all of which result in UV luminosity functions consistent with current high redshift data \citep{McLure_+13,McLeod_16,Livermore_+17,Ishigaki_2017,Atek+18,Oesch_18,Stefanon+19,Bowler+20,Rojas-Ruiz_+20,Bouwens+21,Bouwens+22,Finkelstein_22,Donnan+23, Harikane_2023}. Specifically, the \cutoff, \default, and \boost{} models feature a suppression, a power-law behavior, and an enhancement of star formation efficiency at small halo masses, respectively. We have tuned the escape fraction of ionizing photons in each model so that they achieve similar reionization history, consistent with observations \citep{Ouchi_2010,Mortlock2011,Ono_2012,Schroeder_12,Tilvi_2014, Pentericci_2014,Totani_16,Banados2018,Mason_2018,Hoag_2019,Duruvcikova_20,Jung_2020, Bruton_23}. Additionally, the normalization of the galactic X-ray spectrum varies between the models, resulting in distinct temperature evolution. The \cutoff, \default, and  \boost{} models exhibit a late, moderate, and early rise of the IGM temperature, respectively.  

In Fig.~\ref{fig:properties 3models}, we plot the stellar-to-halo function $f_{\rm *}$ in the left-most panel, the escape fraction $f_{\rm esc}$ in the second panel, the reionization history \xHII(z) in the third panel, and the evolution of the kinetic temperature in the right-most panel, for the \cutoff, \default, and \boost{} models represented in orange, blue, and green colors, respectively. The halo catalogs and dark-matter density fields used in this study were obtained with the $N$-body code, {\tt Pkdgrav3} \citep{Potter2017}, in a 147 cMpc cosmological box, with $2048^{3}$ dark matter particles, resulting in a minimum halo mass of $M_{\rm h, min} = 1.47\times 10^{8} M_{\odot}$. The density fields and halo catalogs are saved every 10 Myr between $z=25$ and 6.

\begin{figure*}
    \centering
    \includegraphics[width =1\textwidth,trim=0.0cm 0.cm 0cm 0.0cm,clip]{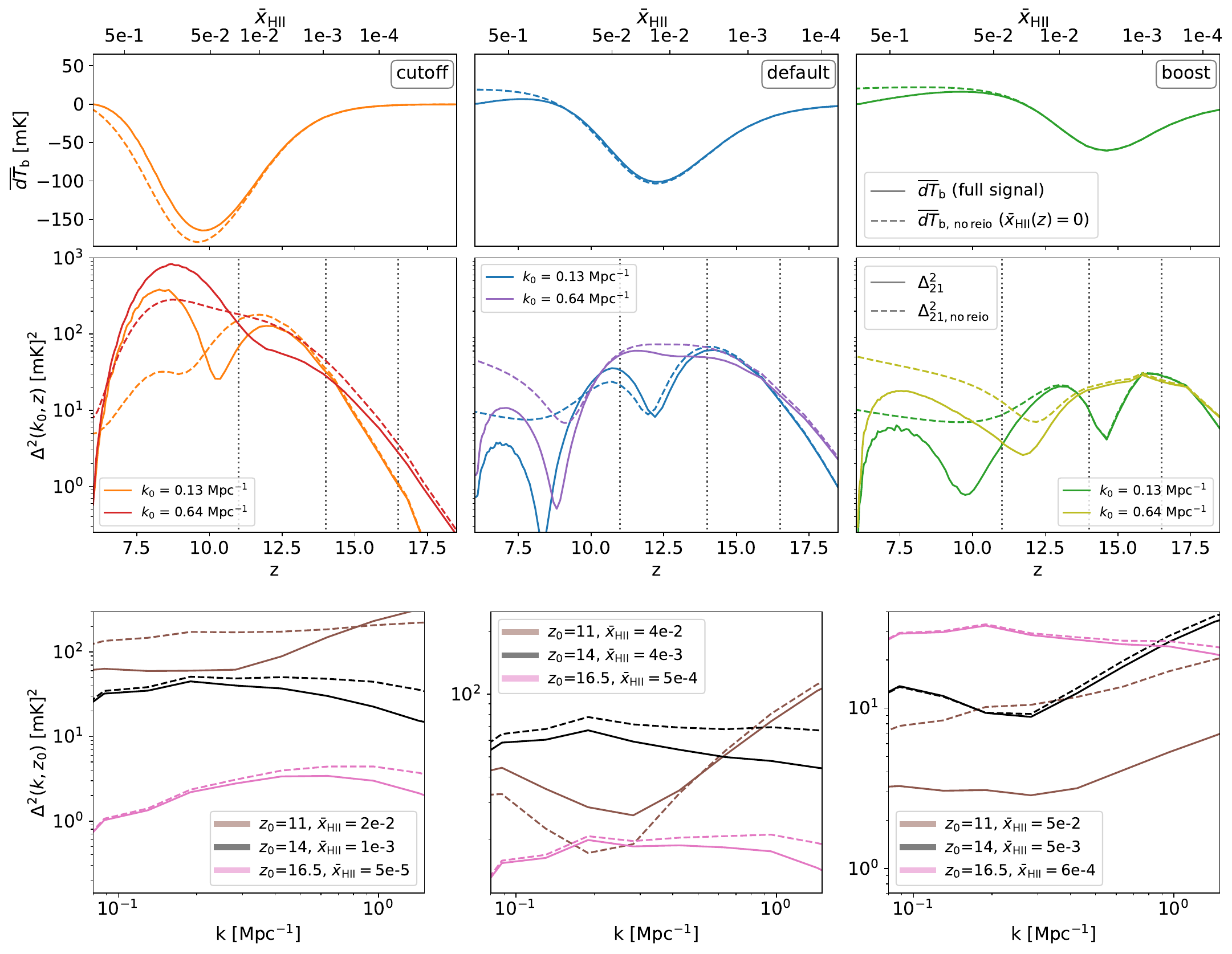}
    \caption{\textbf{Impact of reionization on the 21cm signal}. The \cutoff, \default, and \boost{} astrophysical models are represented in the leftmost, middle, and rightmost columns, respectively. In every panel, solid lines correspond to the full signal including reionization, while dashed lines represent the signal when the universe is assumed to be fully neutral ($\bar{x}_{\rm HII}=0$). \textit{First row}: spatially averaged brightness temperature \dTbbar{}. The relative differences in the global signal are larger or equal to the mean neutral fraction. \textit{Second row}: 21cm power spectrum $\Delta^{2}_{21}$, shown as a function of redshift $z$, at two different scales $k=0.13 \,\text{Mpc}^{-1}$ and $k=0.64\, \text{Mpc}^{-1}$, distinguished by different colors. The differences in $\Delta^{2}_{21}$ appear at earlier epochs compared to the global signal. The dotted vertical lines indicate the redshifts for which we display spectra as a function of scale in the fourth row. 
    \textit{Third row}: 21cm power spectrum $\Delta^{2}_{21}$ as a function of Fourier mode k, at three different redshifts $z=11,\,14,\,\text{and}\,16.5$, where PAPER and MWA have already collected upper limits on the signal \citep{Kolopanis_2019,Yoshiura_2021}. The 21cm power spectrum is extremely sensitive to the presence of rare ionized bubbles.}
    \label{fig:no reio during CD}
\end{figure*}
 
\section{Can we treat separately the EoR and the cosmic dawn?}\label{sec:3}
Three distinct mechanisms govern the evolution of the
21cm signal: the coupling of the spin temperature to
the kinetic temperature induced by \lyal photons, the heating of the gas primarily due to X-ray photons and the
growth and percolation of ionized bubbles produced by
ionizing photons. These processes lead to the characteristic absorption trough in the global signal and the three-peak structure of the large-scale 21cm power spectrum \cite[e.g.][]{mesinger2013signatures,fialkov2014rich,ghara201521,mesinger2016evolution}.

While these three mechanisms typically operate at different epochs, their effects overlap. For instance, rare and small ionized bubbles are already present during the epoch of \lyal{} coupling and heating. Moreover, the universe may not be uniformly heated during the EoR, when ionization fluctuations dominate the signal. \textcolor{black}{This raises questions about the impact of ionization on the cosmic dawn signal and the impact of Lyman-$\alpha$ coupling and heating on the EoR signal.}


A very common approximation in the literature is to separate the signals from the cosmic dawn and the epoch of reionization. Studies focusing on the reionization process often neglect potential fluctuation of the spin temperature \citep{Mellema_2006_simulating,Zahm_11_Comparison_RT_SemNum,Majumdar_14_Comparison_RT_SemNum,Grizzly_comparison_3D_1D,Hassan_2016, Georgiev:2021yvq,Georgiev_2024} to simplify the analysis. A similar trick is often done in studies investigating the cosmic dawn where the reionization bubbles are often neglected \citep{Visbal_2012_noreio,Fialkov_2012_no_reio_bis,Fialkov_2013_noreio,  HaloModel_Paper,ZeusMunoz}.

In what follows, we investigate the validity of treating the epoch of cosmic dawn \textcolor{black}{- defined as the period where Lyman-$\alpha$ coupling and heating occur -} separately from the epoch of reionization. First, we investigate the impact of neglecting reionization and assuming a fully neutral universe during cosmic dawn (Sec.~\ref{sec:no reio during CD}). Then, we examine the saturated spin temperature assumption, which assumes a universe fully heated above the CMB temperature \textcolor{black}{and a spin temperature fully coupled to the gas temperature} during the EoR (Sec.~\ref{sec:no heat during reio}).

\subsection{Ignoring reionization during cosmic dawn \texorpdfstring{$\mathbf{(x_{\rm \mathbf{HII}}=0)}$}{}}\label{sec:no reio during CD}
To investigate the impact of reionization on the 21cm signal, we use our simulation boxes of $\rho$, \xal, \xcl, and \Tk{} to generate a set of brightness temperature boxes where the ionization fraction ($x_{\rm HII}$) is assumed to be uniformly equal to 0. We label them with the subscript ``\textit{no reio}'':
\textcolor{black}{\begin{align}\label{eq:dTb no reio}
    dT_{b, \,\rm no\, reio}(\mathbf{x},z) = T_0(z)&\left[1+\delta_b(\mathbf{x},z)\right]\times \\
    &\left(1-\frac{T_{\rm cmb}(z)}{T_{\rm S}(\mathbf{x},z)}\right).\nonumber
\end{align}}
For the three benchmark models, we measure the global signal and power spectrum from these simulation boxes and compare them to the corresponding fiducial quantities. The results of this analysis are displayed in Fig.~\ref{fig:no reio during CD} for the redshift range $6<z<18.5$. Dashed lines represent the ``\textit{no reio}'' case, while solid lines show the full signal. The three columns correspond to the three models \cutoff, \default, and \boost,  from left to right, respectively. 


In the upper row of Fig.~\ref{fig:no reio during CD}, we plot the global signal $\bar{dT}_{b}$. For all three models, the lines agree at the sub-percent level when \xHIIbar$<0.01$. For \xHIIbar$>0.01$, significant differences start to appear between the global signal predictions with and without ionized bubbles. Notably, we find these differences to be larger than $1-\bar{x}_{\rm HII}$ in all three models. For instance when $\bar{x}_{\rm HII}=0.5$, we observe differences that are larger than $50\%$. This is due to the non-zero correlations between the fields that compose the brightness temperature $dT_{\rm b}$. We will further discuss this issue in Sec.~\ref{sec:comparison-GS}.

In the second row of Fig.~\ref{fig:no reio during CD}, we show the 21cm dimensionless power spectra $\Delta^{2}_{\rm 21}$ as a function of redshift, with and without reionisation. We thereby focus on the two k-modes $k=0.13 \,\text{Mpc}^{-1}$ and $k=0.64\, \text{Mpc}^{-1}$, distinguished by different colours. The third row illustrates $\Delta^{2}_{\rm 21}$ as a function of co-moving Fourier mode $k$ at three different redshifts $z=11, 14$ and 16.5, which roughly cover the heating and Lyman-$\alpha$ dominated regime in our models. They furthermore correspond to the redshift values where we currently have upper limits from PAPER \citep{Kolopanis_2019} and MWA \citep{Yoshiura_2021}. 

Examining the power spectra, we note that the differences between the ``\textit{no reio}'' and the full signal appear at earlier redshifts compared to the global signal. They also vary substantially across the three models. We find that above the threshold \xHIIbar$=0.05$, assuming a fully neutral universe leads to a bias of up to an order of magnitude in the power spectrum. Depending on the source model, this bias manifests as a suppression or enhancement. In the \cutoff{} model, the presence of ionized bubbles enhances the 21cm power spectrum during the EoR, by creating a strong contrast between ionized regions with no signal ($dT_{\rm b}=0$) and cold regions with negative signal ($dT_{\rm b}<0$). In contrast, in both the \default{} and \boost{} models, the IGM is significantly heated when \xHIIbar$>0.05$. Thus, including ionized bubbles in these models decreases the 21cm power spectrum during the EoR by suppressing the peaks of the matter density field. 

Focusing on the regime with \xHIIbar$<0.05$ (where less than five percent of the Universe is ionized), we find that the power spectra still differ by up to a factor of $\sim 3$. \textcolor{black}{Remarkably, these differences are substantially larger than $(1-\bar{x}_{\rm HII})^{2}$, indicating they are primarily to the fluctuations of the $x_{\rm HII}$ field rather than the incorrect global signal.} This is best visible in the bottom row of Fig.~\ref{fig:no reio during CD} where all power spectra are at an ionization fraction below \xHIIbar$=0.05$. The differences between the dashed and solid lines highlight the significant impact of the first ionized bubbles, even if they occupy only a very subdominant fraction of the simulation volume. 

At epochs characterized by \xHIIbar$<0.01$, the influence of reionization on the power spectrum diminishes but remains visible. Overall, neglecting reionization in this regime tends to amplify the power spectrum, as revealed by the pink and black lines in the lower row of Fig.~\ref{fig:no reio during CD}. In the \default{} model, the impact is less pronounced but still noticeable, affecting the power spectrum by up to $10\%$ and $50\%$ at large and small scales, respectively. Finally, in the \boost{} model, the large-scale power spectrum experiences a shift of a few percent, while small scales are impacted by up to $10\%$.

It is worth noting that a fixed ionization fraction has a variable impact on the 21cm power spectrum depending on the astrophysical model. For instance, \xHIIbar{} reaches $10^{-4}$ at $z=16$ in the \cutoff{} model, and at $z=18$ in the \default{} and \boost{} models. As visible from the evolutions of the spin temperature displayed in the right panel of Fig.~\ref{fig:properties 3models}, the \lyal{} coupling is more advanced at these stages in the latter two models compared to the former. The more advanced the UV coupling, the more abundant the absorption regions (\dTb$<0$), and the smaller the effect of the rare ionized bubbles on the power spectrum. Consequently, a larger fraction of regions in absorption is suppressed by ionized pixels in the \cutoff{} model compared to the \boost{} and \default{} models. \textcolor{black}{Overall, the more reionisation and cosmic dawn overlap, the greater the impact of neglecting reionisation on the power spectrum.}

We conclude that neglecting reionization during cosmic dawn when \xHIIbar$<0.01$ may result in an enhancement of the signal, increasingly more pronounced towards small scales. As indicated in the bottom panels of  Fig.~\ref{fig:no reio during CD}, the amplitude of this enhancement factor varies depending on the value of the ionization fraction and the astrophysical parameters. This shows the importance of including the modeling of the reionization process for future analysis focused on the epoch of \lyal{} coupling and heating. Future studies aiming to constrain regions of the parameter space with cosmic dawn upper limits on the 21cm power spectrum may yield biased results if their modeling pipeline overlooks reionization.

\begin{figure*}
    \centering
    \includegraphics[width =1\textwidth,trim=0.0cm 0.cm 0cm 0.0cm,clip]{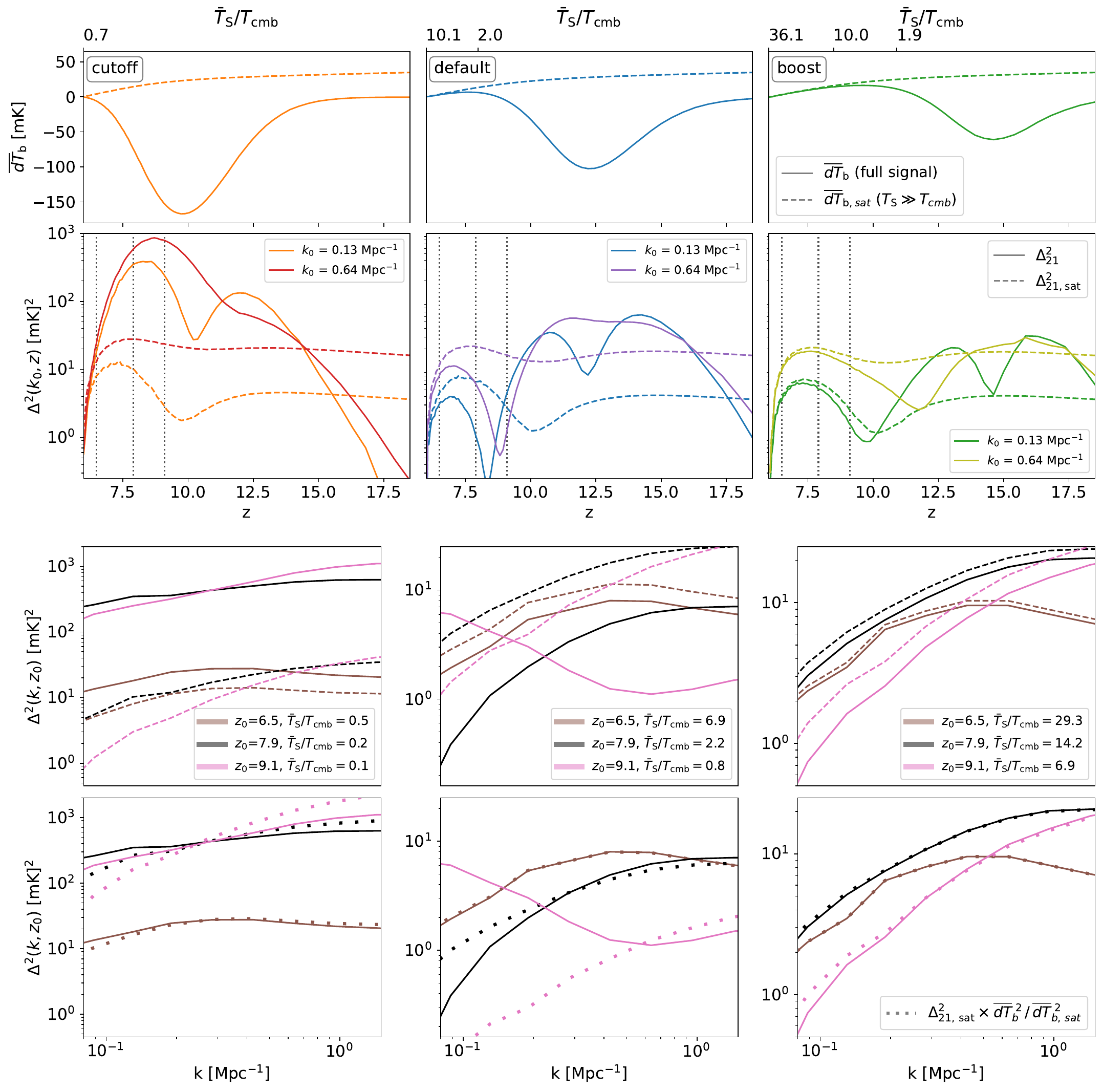}
    \caption{\textbf{Impact of the saturated spin temperature assumption on the 21cm signal.} The \cutoff, \default, and \boost{} models are represented in the leftmost, middle, and rightmost columns, respectively. In every panel, solid lines correspond to the full signal including temperature fluctuations, while dashed lines represent the signal assuming spin temperature saturation ($T_{\rm S}\gg T_{\rm cmb}$). \textit{First row}: spatially averaged brightness temperature \dTbbar{}. We find more than $10\%$ differences once $T_{\rm S}/T_{\rm cmb}$ drops below $10$. \textit{Second row}: 21cm power spectrum $\Delta^{2}_{21}$, plotted as a function of redshift $z$, at two different scales $k=0.13 \,\text{Mpc}^{-1}$ and $k=0.64\, \text{Mpc}^{-1}$, distinguished by different colors. The differences in $\Delta^{2}_{21}$ are due to the differences in global signal \dTbbar{} and to the missing temperature fluctuations in the case where the temperature is assumed to be saturated. The dotted vertical lines indicate the redshifts at which we display spectra as a function of scale in the fourth row. \textit{Third row}: 21cm power spectrum $\Delta^{2}_{21}$ as a function of Fourier mode k, at three different redshifts $z=6.5,\,7.9,\,\text{and}\,9.1$, where upper limits on the signal have been collected by MWA, HERA, and LOFAR, respectively \citep{Trott_MWA_upper_lim,HERA:2021noe, Mertens:2020llj}.\textcolor{black}{\textit{Fourth row}: The dotted line represents the quantity $\Delta^{2}_{\rm 21, \,sat} \times dT_{b}^{2}/dT^{2}_{b,\,sat}$. This corresponds to the signal assuming a non-saturated but homogeneous spin temperature, neglecting the fluctuations of the Lyman-$\alpha$ coupling and kinetic temperature fields.} }
    \label{fig:no heating during reio}
\end{figure*}

\subsection{Ignoring cosmic dawn during reionization \texorpdfstring{$\mathbf{(T_{\rm \mathbf{S}}\gg T_{\rm \mathbf{cmb}})}$}{}}\label{sec:no heat during reio}
We now focus on the epoch of reionization (EoR), investigating the effect of temperature and Lyman-$\alpha$ fluctuations on the signal. A common approximation consists of assuming a saturated spin temperature $T_{\rm S}\gg T_{\rm cmb}$ during the whole EoR period. \textcolor{black}{This situation occurs when Lyman-$\alpha$ coupling is saturated ($x_{\rm \alpha}\gg 1$) and when the IGM is fully heated well above the CMB temperature  ($T_{\rm k}\gg T_{\rm cmb}$). 
As in the previous subsection, we generate a set of boxes labelled $dT_{b,\,sat}$ (standing for ``\textit{saturated}'') defined by 
\begin{align}\label{eq:dTb sat}
    dT_{b, \,sat}(\mathbf{x},z) = T_0(z) x_{\rm HI}(\mathbf{x},z)\left[1+\delta_b(\mathbf{x},z)\right].
\end{align}}
Then, we compute the power spectrum and global signal from these boxes, which we compare to the true signal. Our findings are displayed in Fig.~\ref{fig:no heating during reio}. The three columns again correspond to the three models, \cutoff, \default, and \boost{} (from left to right). The dashed and solid lines represent the signals with and without saturated spin temperature, respectively. 

In the top row of Fig.~\ref{fig:no heating during reio} we plot the global signal for the two cases. Not surprisingly, the saturated case only agrees with the full calculation when the spin temperature becomes significantly larger than the CMB temperature (see top axis showing the temperature ratio). Below $\bar{T}_{\rm S}/T_{\rm cmb}\sim 10$ differences start to become visible and below $\bar{T}_{\rm S}/T_{\rm cmb}\sim 2$ the two curves start to deviate strongly.

Similar conclusions can be drawn from the second row of Fig.~\ref{fig:no heating during reio}, where we depict the 21cm power spectra $\Delta^{2}_{\rm 21}$ as a function of redshift, at two different scales $k=0.13 \,\text{Mpc}^{-1}$ and $k=0.64 \,\text{Mpc}^{-1}$. There is no well-delimited period in redshift where the saturated case yields results in agreement with the full calculation. Only in the \boost{} model are the two lines close together for redshifts below 10. However, a closer inspection still reveals differences between 5 and 20 percent. For the \default{} and the \cutoff{} model, the ``\textit{saturated}'' and true results are substantially different, even at late stages of reionization.

In the third row of Fig.~\ref{fig:no heating during reio}, we display $\Delta^{2}_{\rm 21}$ as a function of comoving Fourier mode $k$ at three different redshifts $z=6.5, 7.9$ and 9.1. These redshifts are selected to be in the regime where reionization is believed to have occurred. They correspond to the redshift values from the current upper limits of MWA \citep{Trott_MWA_upper_lim}, HERA \citep{HERA:2021noe} and LOFAR \citep{Mertens:2020llj}, respectively. The differences between the saturated case and the full calculation lie between about 10 percent in the best case and 2-3 orders of magnitude in the worst.


The \cutoff{} model corresponds to an example of cold reionization where $\bar{T}_{\rm S}/T_{\rm cmb}$ remains below 1 across all redshifts. Therefore, the assumption of a saturated spin temperature is trivially not fulfilled and, hence, there is a very strong disagreement between the saturated and the full cases in Fig.~\ref{fig:no heating during reio}. 
\textcolor{black}{Although the \cutoff{} model may seem extreme, it is worth noting that it remains a valid scenario not ruled out by any current observation. The power spectrum of the \cutoff{} model is below all the available upper limits, and at $z=7.92$, the average spin temperature is  $\bar{T}_{\rm S}(z=7.92)=4.25\rm \,K$, remaining above the current lower limit of $T_{\rm S}=2.3\rm \,K$ obtained by Ref.~\citep{HERA:2021noe} within 95$\%$ confidence interval.}

In the \default{} and \boost{} models, more efficient X-ray heating drive the spin temperature above $T_{\rm cmb}$ before $z=6$. In these two models, the saturated and full global signal calculations agree to better than $10\%$ once the fraction $\bar{T}_{\rm S}/T_{\rm cmb}$ goes above 10. In the power spectrum, on the other hand, differences of order $10\%$ remain as long as $\bar{T}_{\rm S}/T_{\rm cmb}<100$. At $\bar{T}_{\rm S}/T_{\rm cmb}\simeq 10$ the errors due to the saturated spin temperature assumption is typically between 20 and 50 percent.

\textcolor{black}{The difference in the power spectra between the saturated and full signals is due to both the incorrect global signal and the missing spin temperature fluctuations. To isolate the effect of neglecting spin temperature fluctuations while maintaining the correct global signal, we plot the quantity $\Delta^{2}_{\rm 21, ,sat} \times dT_{b}^{2}/dT^{2}_{b,,sat}$ as dotted lines in the fourth row of Fig.~\ref{fig:no heating during reio}. We achieve per cent level agreement with the true signal when $\bar{T}_{\rm S}/T_{\rm cmb}>10$, indicating most of the error arises from the incorrect global signal in this case. However, substantial discrepancies remain when $\bar{T}_{\rm S}/T_{\rm cmb}<10$, showing that temperature fluctuations cannot be ignored in this regime.}

In summary, our analysis shows that a true saturation of the spin temperature is only reached at $\bar{T}_{\rm S}/T_{\rm cmb}\sim 100$ and above. Below this ratio, neglecting fluctuations in the temperature and the Lyman-$\alpha$ coupling yields errors of 10\% or more on the 21cm power spectrum (a number that is strongly rising towards smaller values of $\bar{T}_{\rm S}/T_{\rm cmb}$). For our three benchmark models, the condition of $\bar{T}_{\rm S}/T_{\rm cmb}>100$ is never fulfilled during the EoR epoch. Although other models may have a regime where $\bar{T}_{\rm S}/T_{\rm cmb}$ rises above 100, we cannot know if such a model is realised in nature before we measure the signal. We therefore conclude that assuming a saturated spin temperature is not an adequate strategy for predicting the 21cm signal.

\section{Testing the perturbative approach}\label{sec:testing PT}
To bypass the computational cost of grid-based methods, and efficiently explore the vast astrophysical and cosmological parameter space, analytical techniques have been developed to compute the 21cm global signal and power spectrum within a matter of seconds, without modeling the full cosmological fluctuations of \dTb{} on a grid \citep{Barkana:2004zy,Barkana:2005b,Furlanetto:2006jb,Pritchard:2006sq, Mirocha:2014faa,Raste_2018, HaloModel_Paper, ZeusMunoz,Mirocha:2022pys,HMreio}. These methods are rooted in a perturbative treatment of the \dTb{} field, and should not be confused with analytical methods based on an effective bias expansion of the signal \cite{mcquinn2018observable,qin2022effective}.

In this section, we investigate the accuracy of the perturbative approach for computing the 21cm signal. First, we describe the building blocks of the approach in Sec.~\ref{sec:Perturbative approaches}.  Then, we use our simulation boxes to reproduce the predictions of the perturbative approach, and compare them with the actual signal. We perform this test for our three benchmark models, examining the global signal in Sec.~\ref{sec:comparison-GS} before turning to the power spectrum in Sec.~\ref{sec:comparison PS}.

\subsection{The perturbative approach for 21cm}\label{sec:Perturbative approaches}

The core idea of the perturbative approach for the 21cm signal involves expressing the \dTb{} perturbation as a sum and product of individual perturbations arising from the matter, the ionization fraction, the kinetic temperature, and the \lyal{} coupling coefficient fields. This decomposition is achieved through a Taylor expansion (as detailed in Sec.~\ref{sec:Taylor exp}), and by neglecting high-order products of these fields (as described in Sec.~\ref{sec:linearity}). Subsequently, the 21cm power spectrum is obtained as a sum of auto and cross power spectra of these individual fields (see Sec.~\ref{sec:final expression}). These spectra can then be computed using various methods.

\subsubsection{Decomposition of \texorpdfstring{$dT_{\rm b}$}{} into individual components}\label{sec:decomposition}

The fluctuations of \dTb{} are sourced by four space and time-dependent fields: \dr, $\delta_{\rm U}$, $\delta_{V}$ and \db, representing the fractional perturbations of the \xHII, $U_{\rm \alpha}$, $V_{\rm k}$, and the matter field, respectively. Accordingly, Eq.~(\ref{eq:dTb}) can be reformulated as follows:
\begin{equation}
  dT_{\rm b}(\mathbf{x},z) = \widehat{dT_{\rm b}}(1+\beta_b\delta_{b})(1+\beta_r\delta_{r})(1+\delta_{U})\\(1+\delta_{V}),
  \label{eq:dTb perturb 1}
\end{equation}
with
\begin{equation}
  \widehat{dT_{\rm b}}(z) =T_0(z) \bar{x}_{\rm HI}(z) \bar{U}_{\rm \alpha}(z)\bar{V}_{\rm k}(z),
  \label{eq:mean dTb 1}
\end{equation}
where the horizontal bars above letters designate spatially averaged quantities and where the $\beta$-factors are given by $\beta_r=-(1-{\bar x}_{\rm HI})/{\bar x}_{\rm HI}$, $\beta_b= 1$ and only depend on redshift. Note that there is no assumption underlying Eq.~(\ref{eq:dTb perturb 1}), the equality is exact.

\subsubsection{The Taylor expansion approximation}\label{sec:Taylor exp}
To move forward and obtain a simpler expression for \dTb{}, the $U_{\rm \alpha}$ and $V_{\rm k}$ fields are replaced by their \textit{Taylor series} truncated at order 1. One then obtains
\begin{equation}\label{eq:U expansion}
    U_{\rm \alpha,\,taylor} = \frac{\bar{x}_{\rm tot}}{1+\bar{x}_{\rm tot}}(1+\beta_{\alpha}\delta_{\alpha}),\\
\end{equation}
\begin{equation}\label{eq:V expansion}
    V_{\rm k,\,taylor} = \left[1-\frac{T_{\rm cmb}}{\bar{T}_{\rm k}}\right](1+\beta_T\delta_{T}),\\
\end{equation}
with $\beta_{\alpha}={\bar x}_{\alpha}/{\bar x}_{\rm tot}/(1+{\bar x}_{\rm tot})$ and $\beta_T=T_{\rm cmb}/({\bar T}_{\rm k}-T_{\rm cmb})$. Note that Eq.~(\ref{eq:U expansion}) and Eq.~(\ref{eq:V expansion}) are valid whenever
\begin{equation}\label{eq:validity Taylor}
\begin{aligned}
\delta^{*}_{\rm \alpha}\equiv  \frac{\bar{x}_{\rm \alpha}}{1+\bar{x}_{\rm tot}}\delta_{\rm \alpha}\ll 1 \quad \text{and}  \quad 
\delta_{\rm T}\ll 1 \ .
\end{aligned}
\end{equation}
The Taylor expansion transforms the expression of \dTb{} into
\begin{multline}\label{eq:dTb nl taylor}
  dT_{\rm b,taylor} = \widetilde{dT_{\rm b}}(1+\beta_b\delta_{b})(1+\beta_r\delta_{r})\times\\(1+\beta_T\delta_{T})(1+\beta_{\alpha}\delta_{\alpha}) \ ,
\end{multline}
with
\begin{equation}
  \widetilde{dT_{\rm b}} = T_0(z) \bar{x}_{\rm HI}(z) \frac{\bar{x}_{\rm tot}(z)}{1+\bar{x}_{\rm tot}(z)} \left[1-\frac{T_{\rm cmb}(z)}{\bar{T}_{\rm k}(z)}\right] \ . \\ 
  \label{eq: fake dTb}
\end{equation}
Eq.~(\ref{eq:dTb nl taylor}) corresponds to a multivariate polynomial function of the 4 individual perturbation fields. It is the starting point to compute $P_{\rm 21}$ perturbatively. Note that $dT_{\rm b,taylor}$ being an approximation of the true \dTb{} field, it may lead to inaccurate results, especially near the center of halos where \xal{} and \Tk{} may deviate significantly from their mean values, rendering the conditions of Eq.~(\ref{eq:validity Taylor}) invalid. Sec.~\ref{sec:comparison PS} will further explore this issue.


\subsubsection{Linearity: neglecting high-order  perturbations in \texorpdfstring{$\delta_{\rm \alpha}$}{}, \texorpdfstring{$\delta_{\rm T}$}{}, and \texorpdfstring{$\delta_{\rm b}$}{}}\label{sec:C.3}\label{sec:linearity}
Eq.~(\ref{eq:dTb nl taylor}) contains 16 individual products of fluctuations, which would lead to 16 + $\binom{16}{2}$= 136 auto and cross terms when computing the power spectrum of \dTb{}. Therefore, it is critical to \textit{a priori} neglect some of these terms. Since by construction \dr{} is of order $\mathcal{O}(1)$, the perturbations in \dr{} must be kept to non-linear order. 
In previous studies such as \citep{Pritchard:2006sq, HMreio}, every term including more than two perturbations in either \db, \da, or \dT{} was discarded. This assumption was supported quantitatively by measuring the standard deviations of these three fields, which remain below one. 
This approach yields a more manageable expression for \dTb{}, which we indicate with the subscript ``\textit{nl,r}'':
\begin{multline}\label{eq:dTb nl, r}
dT_{\rm b}^{(nl,\,r)} = \widetilde{dT_{\rm b}}(1+\beta_r\delta_{r} + \beta_b\delta_{b} + \beta_T\delta_{T} + \beta_{\alpha}\delta_{\alpha}+ \\ \beta_r\beta_b\delta_r\delta_b + \beta_r\beta_T\delta_r\delta_T + \beta_r\beta_{\alpha}\delta_r\delta_{\alpha}).
\end{multline}


\subsubsection{Final perturbative expression for the power spectrum}\label{sec:final expression}
 
Moving forward with Eq.~(\ref{eq:dTb nl, r}), we obtain the final decomposition for the 21cm power spectrum:
\begin{equation}\label{eq:P21 decomp}
P_{21, decomp} = P^{\rm (lin)}_{21} + P^{\rm (nl, r, \,1)}_{21}+ P^{\rm (nl, r, \,2)}_{21} 
\end{equation}
with 
\begin{multline}\label{eq:Plin}
P^{\rm (lin)}_{21} = \widetilde{dT_{\rm b}}^{2}\times\left[ P_{r,r} + P_{b,b} + P_{T,T} + P_{\alpha,\alpha}\right. \\
 + 2\left(P_{r,b} + P_{r,T} + P_{r,\alpha} + P_{b,T} + P_{b,\alpha} + P_{T,\alpha}\right) \left. \vphantom{P_{r,m}}\right],  
\end{multline}
\begin{multline}\label{eq:Pnl,r,1}
P^{\rm (nl,r, \,1)}_{21} =  \widetilde{dT_{\rm b}}^{2}\times[2( P_{r,rb} +P_{b,rb} )+P_{rb,rb}], 
\end{multline}
\begin{multline}\label{eq:Pnl,r,2}
P^{\rm (nl,r, \,2)}_{21} =  \widetilde{dT_{\rm b}}^{2}\times\left[2( P_{r,rT} + P_{r,r\alpha}\right. \\
 + P_{r,b\alpha} + P_{r,bT} + P_{r,aT} + P_{b,rT} + P_{b,r\alpha}  \\
+P_{T,rb} + P_{T,rT} + P_{T,r\alpha} +P_{\alpha,rb} + P_{\alpha,rT} + P_{\alpha,r\alpha}\\
+P_{rb,rT} + P_{rb,r\alpha} + P_{rT,r\alpha}+ P_{rb\alpha,r} + P_{rT\alpha,r} + P_{rbT,r}) \\ + P_{rT,rT} + P_{r\alpha,r\alpha}\left.\vphantom{P_{r,m}}\right] \ .
\end{multline}
We isolated the higher-order contributions arising from matter (\db) and ionization (\dr) perturbations, into $P^{\rm (nl,r, \,1)}_{21}$ for convenience. Note that all the individual terms above are power spectra of perturbative quantities \dr, \db, \da, \dT, including the $\beta$ pre-factors. For instance, $P_{\alpha, rT}$ is defined as the Fourier transform of $\beta_{\rm \alpha}\beta_{\rm r}\beta_{\rm T}\big<\delta_{\alpha}(x) \delta_{r}(x')\delta_{\rm T}(x')\big>$. We refer to \citep{Lidz:2006vj} for a more detailed study of these higher-order terms.
The last three equations are the building blocks of the analytical approach for the 21cm power spectrum. We will test their validity in Sec.~\ref{sec:comparison PS}.

\begin{figure*}
    \centering
    \includegraphics[width =1\textwidth,trim=0.0cm 0.cm 0cm 0.0cm,clip]{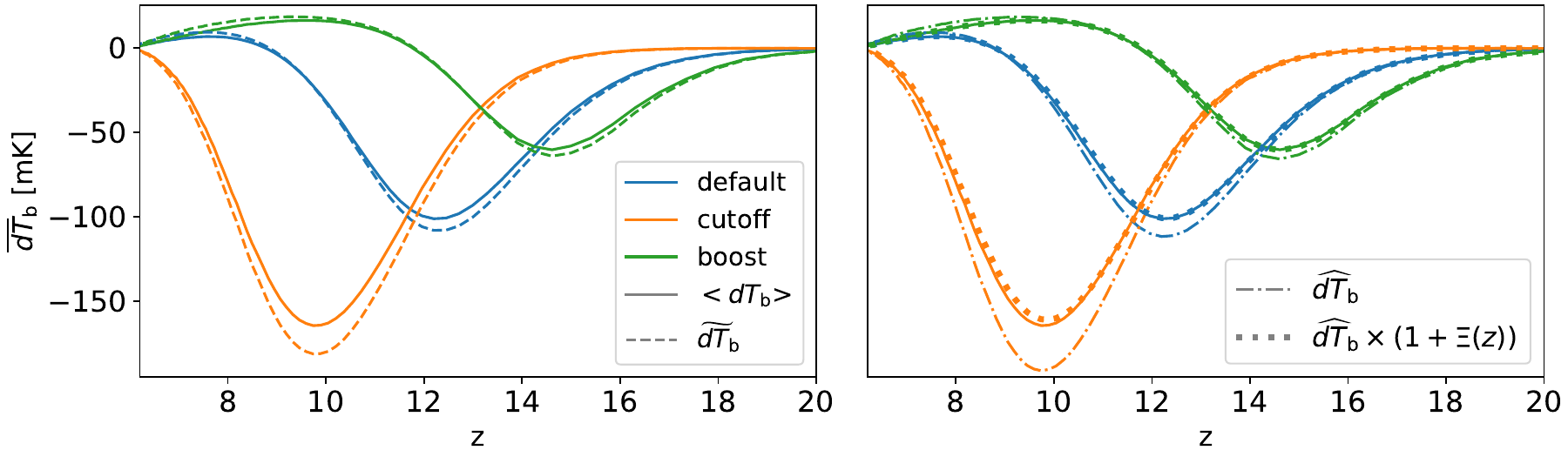}
    \caption{\textbf{Impact of calculating the 21cm global signal from the mean quantities of the ionization fraction (\xHIIbar{}), the temperature \Tkbar{} and the Lyman-$\alpha$ coupling coefficient \xalbar{}}. The three models \default{}, \cutoff{}, and \boost{}, are represented in blue, orange, and green, respectively. In both panels, solid lines correspond to the ``true'' global signal, computed as the spatial average of the \dTb(x,z) simulation boxes. In the \textit{left panel}, the dashed line represents  $\widetilde{dT_{\rm b}}$, defined in Eq.~(\ref{eq: fake dTb}) and computed via the mean individual quantities \Tkbar, \xalbar, and \xHIIbar. We observe a difference of the order of 10 mK around the dip of the absorption trough. In the \textit{right panel}, the dashed-dotted line corresponds to $\widehat{dT_{\rm b}}$, defined in Eq.~(\ref{eq:mean dTb 1}) and computed via the mean individual quantities $\bar{V}_{\rm k}$, $\bar{U}_{\rm \alpha}$, and \xHIIbar{}. We observe an even larger difference with the true signal. To recover the global signal with percent precision, it is sufficient to multiply $\widehat{dT_{\rm b}}$  with the correction factor $(1+\Xi(z))$, which involves calculating the 6 auto and co-variances between the fields \U{}, \V{}, $\rho$, and \xHII{}.}
    \label{fig:GS 3models}
\end{figure*}

\subsection{Global Signal}\label{sec:comparison-GS}
The spatially averaged brightness temperature \dTbbar{} can be used to constrain both the cosmological and the astrophysical parameters. It is experimentally very challenging to measure, but several single-dish experiments currently attempt to detect the global 21cm signal during cosmic dawn and reionization \citep{BigHorns, SARAS_2, REACH,PRIZM, LEDA,MIST}. The EDGES detection \citep{Edges_2018}, although being highly debated and even excluded at $95\%$ confidence by the SARAS3 experiment \citep{SARAS3_excludingEdges},  triggered the community to explore the rich constraints that can be extracted from the global signal \citep[see e.g.][]{Mirocha:2018cih,Amico_18,Fialkov_18,Hektor_18,Brandenberger_2019,Mitridate_2018,Wang_2018,Mittal:2020}. Therefore, it is crucial to have at our disposal reliable and computationally efficient tools to predict the global signal.

Analytical codes such as {\tt ARES}\citep{Mirocha:2014faa}, {\tt HMreio}\citep{HMreio} or {\tt ZEUS}\citep{ZeusMunoz} can predict the global 21cm signal in a mere second. They compute the global signal \dTbbar{} based on the average quantity of the mean ionization fraction \xHIIbar, the mean temperature of the gas \Tkbar, and the average \lyal{} coupling coefficient \xalbar. Therefore, they determine the quantity $\widetilde{dT_{\rm b}}$, as defined in Eq.~\ref{eq: fake dTb}. These methods implicitly assume two things: (i) that the mean of the product of several fields equals the product of their means, such that \dTbbar$=\widehat{dT_{\rm b}}$ (Eq.~\ref{eq:dTb perturb 1}), and (ii) that the mean of the fields \U{} and \V{} can be computed via the mean quantities \xalbar{} and \Tkbar{}, such that $\widehat{dT_{\rm b}}=\widetilde{dT}_{\rm b}$. Combining these two assertions leads to \dTbbar$=\widetilde{dT_{\rm b}}$.

In the left-hand panel of Fig.~\ref{fig:GS 3models}, we display $\widetilde{dT_{\rm b}}$, and \dTbbar{}, computed from our simulation maps, as dashed and solid lines. The \default, \cutoff, and \boost{} models are represented in blue, orange, and green colors. We observe a noticeable discrepancy between these two quantities. Across all the models, we find a relative error of about 10 $\%$ around the dip of the absorption trough. In the EoR period (where the signal may be observed in emission) the relative error can reach a factor of 2. In terms of absolute error (expressed in mK) the largest values are found around the dip of the absorption trough, with values of 15, 10, and 5 mK for the \cutoff, \default, and \boost{} models, respectively. 

In the right-hand panel of Fig.~\ref{fig:GS 3models}, we show the quantity $\widehat{dT}_{\rm b}$ as dash-dotted line. $\widehat{dT}_{\rm b}$ is computed in a similar way than $\widetilde{dT_{\rm b}}$, but using the mean of the individual quantities \U{} and \V{} instead of \Tk{} and \xal{}. Compared to $\widetilde{dT_{\rm b}}$, we find an even more pronounced discrepancy of the order of $15\%$. The difference between $\widehat{dT}_{\rm b}$ and $\widetilde{dT_{\rm b}}$ is because $\bar{U}_{\rm \alpha}\neq \bar{x}_{\rm \alpha}/(1+\bar{x}_{\rm \alpha})$, and $\bar{V}_{\rm k}\neq (1- T_{\rm cmb}/\bar{T}_{\rm k})$, while the difference between $\widehat{dT}_{\rm b}$ and \dTbbar{} arises from the fact that the mean of the product of multiple fields is not the product of their means. \textcolor{black}{These two approximation steps might have opposite effects. For example, in our three models, $|\bar{V}_{\rm k}|> |(1- T_{\rm cmb}/\bar{T}_{\rm k})|$ around the dip of the absorption trough, causing  $\widetilde{dT_{\rm b}}$ to be more accurate than $\widehat{dT}_{\rm b}$, despite relying on one more degree of approximation.}



We can understand the connection between $\widehat{dT_{\rm b}}$ and the actual global signal \dTbbar{} by taking the spatial average of equation Eq.~(\ref{eq:dTb perturb 1}). Ignoring all terms with more than 2 delta-terms we obtain
\begin{equation}\label{eq:mean dTb nl taylor}
  \langle dT_{\rm b} \rangle  \simeq \widehat{dT_{\rm b}}(1+\Xi(z))
\end{equation}
with 
\begin{equation}\label{eq:Xi factor}
  \Xi(z) = \sum_{\substack{i, j \in \{r, b, V_{\rm k}, U_{\rm \alpha}\}\\ i \neq j}} \sigma_{ ij}(z),
\end{equation}
where $\sigma_{ ij}(z)= \langle\delta_{i}\delta_{j}\rangle$ denotes the covariance of the two fields $\delta_{i}$, and $\delta_{j}$. We calculate $\sigma_{ij}(z)$ directly from our simulation boxes, as the spatial average of the product of two fields. $\Xi$ depends only on redshift and is a sum of 6 different terms. In the right panel of Fig.~\ref{fig:GS 3models}, we plot the quantity $\widehat{dT_{\rm b}}(1+\Xi(z))$. We observe that it provides a very good approximation of the true global signal, reducing the relative error around the dip of the absorption trough to below $2\%$. 


In summary, we find that analytical techniques for computing the 21cm global signal from the average of the individual fields lead to a $\sim 10\%$ error on the signal around the dip of the absorption trough. This error is primarily due to neglecting non-zero correlations between the fields. \textcolor{black}{However, we find that three-point and higher correlations have only a percent-level impact on the global signal and can be ignored. Therefore, accurate results with analytical models are achievable, provided they can calculate the covariance between the individual fields. }

\subsection{Power spectrum}\label{sec:comparison PS}
In this section, we test and discuss the validity of the perturbative approach to compute the 21cm power spectrum based on the decomposition of \dTb{} into individual components (Eq.~\ref{eq:P21 decomp}), and described in Sec.~\ref{sec:Perturbative approaches}. Using the individual \xHII, \xal, \Tk, \db{} boxes for our three astrophysical models, we compute the various auto and cross power spectra appearing in Eq.~(\ref{eq:Plin},\ref{eq:Pnl,r,1},\ref{eq:Pnl,r,2}). This allows us to estimate the relevance of each term and examine the accuracy of the perturbative approach to model the 21cm power spectrum.

We begin by testing the performance of the linear prediction $P_{\rm 21}=P_{\rm 21}^{\rm lin}$, defined in Eq.~(\ref{eq:Plin}). This first approximation assumes full linearity of all the individual fields \dr, \da, \dT, \db. Note that here, "linearity" refers to the assumption that perturbations are on average small, allowing us to neglect products of more than two terms in the \dTb{} decomposition. Under this assumption, \dTb{} reduces to Eq.~(\ref{eq:dTb nl, r}) without the last three higher-order terms involving the \dr{} perturbation. This approximation is known to be inaccurate during the EoR, due to the non-linearity of the ionization fraction field \citep{Lidz:2006vj,Georgiev:2021yvq}, but has not been properly tested during cosmic dawn, and serves as the starting expression upon which we will add corrections.

\begin{figure*}
    \centering
    \includegraphics[width =1\textwidth,trim=0.0cm 0.cm 0cm 0.0cm,clip]{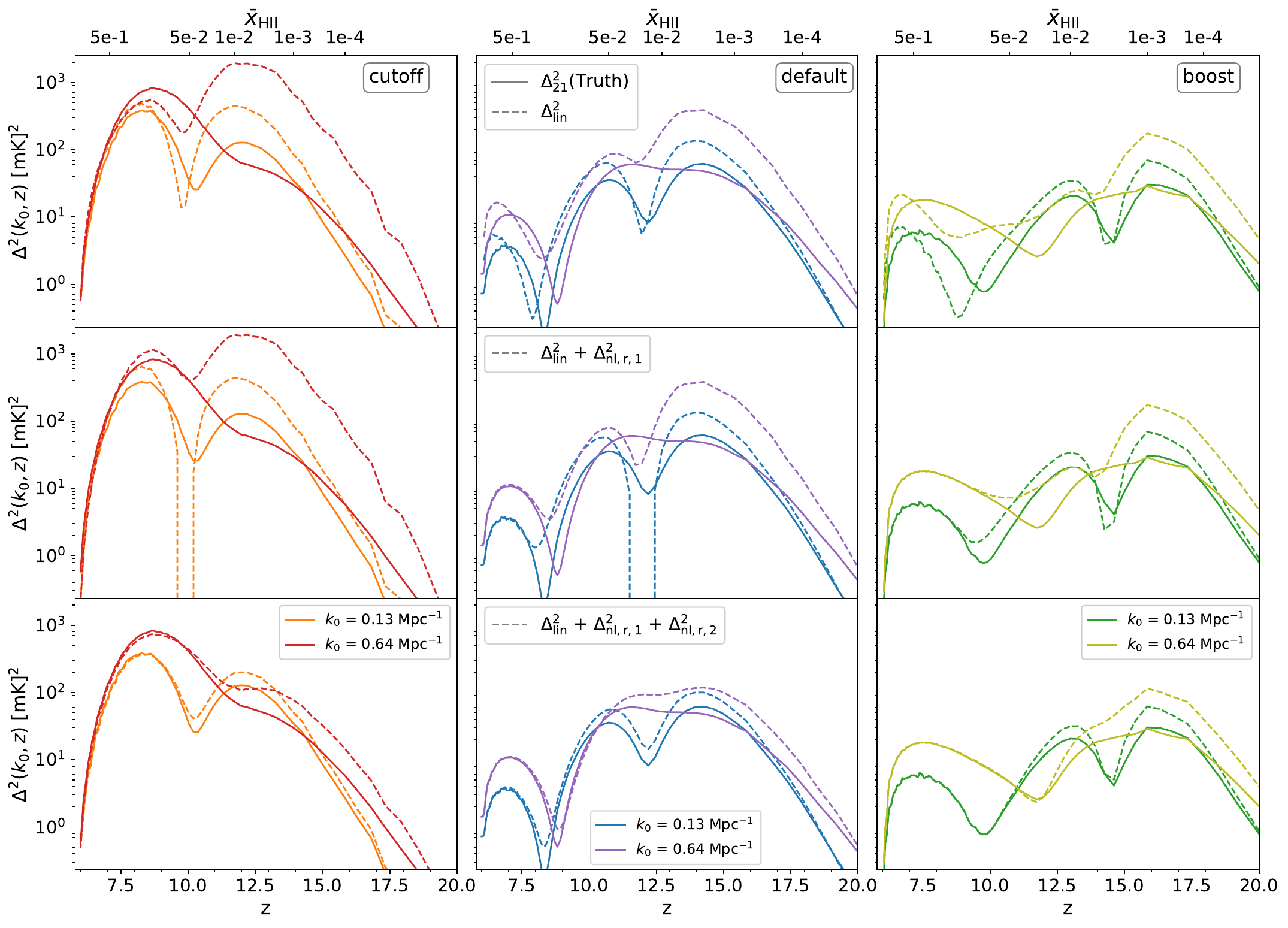}
    \caption{\textbf{Testing the 21cm power spectrum calculation from the perturbative approach.} The three columns correspond to the three benchmark models \cutoff, \default, and \boost, from left to right, respectively. In each panel, we show the 21cm dimensionless power spectrum as a function of redshift, at two different scales k = 0.13, 0.64 Mpc$^{-1}$, distinguished by different colors. The solid lines represent the "true" 21cm power spectrum $\Delta^{2}_{\rm 21}$, measured directly from the \dTb(x,z) boxes. The dashed lines represent various predictions obtained from the decomposition of \dTb{} into individual components. \textit{Upper row}: dashed lines correspond to $\Delta^{2}_{\rm lin}$ (Eq.~\ref{eq:Plin}). We observe strong differences between $\Delta^{2}_{\rm lin}$ and the true signal across the three models, over most scales and redshifts.  \textit{Middle row}: dashed lines represent $\Delta^{2}_{\rm lin}+\Delta^{2}_{\rm nl, r, 1}$ (Eq.~\ref{eq:Pnl,r,1}). The agreement is now remarkable during the EoR for the \default{} and \boost{} models only. \textit{Lower row}: dashed lines correspond to $\Delta^{2}_{\rm lin}+\Delta^{2}_{\rm nl, r, 1}+\Delta^{2}_{\rm nl, r, 2}$. This last expression contains all 24 higher-order contributions from the ionization fraction field \dr{} (Eq.~\ref{eq:Pnl,r,1},\ref{eq:Pnl,r,2}). The agreement with the true signal is now excellent during the EoR for the three models. However, significant discrepancies persist during cosmic dawn.}
    \label{fig:PS decomposition 3 models}
\end{figure*}

Using our simulated boxes, we compute the 10 terms present in Eq.~(\ref{eq:Plin}) and subsequently calculate $P_{\rm 21}^{\rm lin}$. In the three upper panels of Fig.~\ref{fig:PS decomposition 3 models}, we plot the true dimensionless 21cm power spectrum $\Delta^{2}_{\rm 21}$ computed from the \dTb{}-maps (solid lines) next to the linear prediction $\Delta^{2}_{\rm lin}$ (dashed lines). The spectra are displayed at two different scales k = 0.13 and 0.64 Mpc$^{-1}$. The \cutoff, \default,  and \boost{} models are represented in the leftmost, central, and rightmost panels, respectively. 

For the three models, the linear approximation performs poorly over the whole range of redshifts. As expected, the match is better at the largest scale k=0.13 Mpc$^{-1}$. Overall, the linear theory tends to overpredict the clustering signal. This overestimation is of the order of factor 2-3 for the larger scales (k=0.13 Mpc$^{-1}$) going up to a factor 5-10 for the smaller scales (k=0.64 Mpc$^{-1}$).

In the following, we will comment in more detail on the differences visible during both the cosmic dawn and the EoR. We will furthermore investigate to what extent higher-order terms can improve the result.

\subsubsection{The epoch of reionization \texorpdfstring{$\mathit{(z<10)}$}{}}\label{sec:comparison}
The EoR manifests itself as a peak in the 21cm power spectrum, visible between redshift $z=6$ and $z=10$ in our three models. The failure of linear theory is well expected in this regime because the \dr{} field is known to have fluctuations of order unity. This issue was already addressed in the literature \citep{Lidz:2006vj, Georgiev:2021yvq} and solved by including higher-order products in \dr{} to the \dTb{} decomposition. However, these studies investigated the contribution from higher-order terms assuming a saturated spin-temperature, thereby ignoring higher-order contributions arising from the temperature fluctuations. We perform the same kind of analysis but including temperature fluctuations. Namely, we check if the inclusion of the higher-order contributions in the matter (\db) and reionization (\dr) fields (contained in Eq.~\ref{eq:Pnl,r,1}) suffice to recover the true 21cm power spectrum during reionization.

Using our boxes of \xHII{} and \db, we compute the three extra terms contained in Eq.~(\ref{eq:Pnl,r,1}), and add this correction to the linear prediction. The result (denoted as $\Delta^{2}_{\rm lin} + \Delta^{2}_{\rm nl, r, 1}$) is shown in the second row of Fig.~\ref{fig:PS decomposition 3 models}. For  the \default{} and \boost{} models a clear improvement can be observed. With the $\Delta_{\rm nl, r, 1}$ correction, the reionization peak is now accurately recovered in the both models. In the \cutoff{} model, on the other hand, the higher order terms in \dr{} and \db{} do not improve the fit with respect to the linear case. This is caused by the fact that the reionization and temperature peaks are merged in this model, strongly suggesting that at least the temperature fluctuations remain important until the late stages of reionization.

In the three bottom panels of Fig.~\ref{fig:PS decomposition 3 models}, we show $\Delta^{2}_{21, decomp} = \Delta^{2}_{\rm lin} + \Delta^{2}_{\rm nl, r, 1} + \Delta^{2}_{\rm nl, r, 2}$, the 21cm power spectrum consisting of all linear terms plus all the 24 higher-order contributions arising from \dr{} perturbations, as defined in  Eqs.~(\ref{eq:Pnl,r,1},~\ref{eq:Pnl,r,2}). The result of this decomposition is now in very good agreement with the true signal. Not only the \default{} and \boost{} but also the \cutoff{} model now show a reasonable match with respect to the true result, at least for the reionization epoch at $z\lesssim 10$.  



\begin{figure*}
    \centering
    \includegraphics[width =1\textwidth,trim=0.0cm 0.cm 0cm 0.0cm,clip]{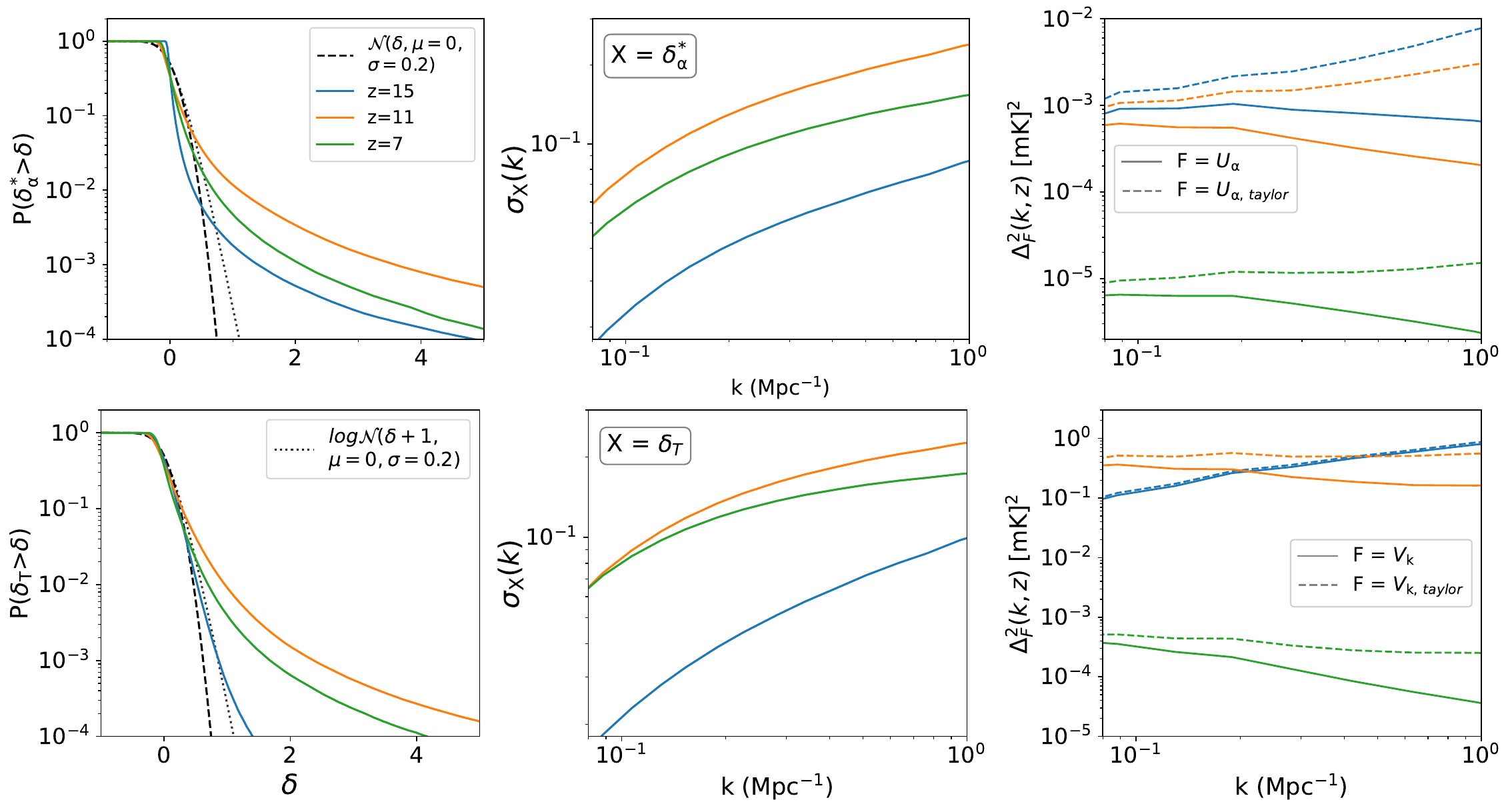}
    \caption{\textbf{Break-down of perturbation theory due to highly non-Gaussian distributions of the \xal{} and \Tk{} fields.} The data displayed in this figure corresponds to the \default{} model. 
    The different colors highlight different redshifts, 
    $z=15,\,11,\,\text{and}\, 7$,  in blue, orange, and green, respectively. 
    \textit{Leftmost column}: Cumulative distribution functions of \dastar{} and of \dT{} in the upper and lower panels, respectively. The black dashed and dotted lines represent the distribution functions of a Gaussian and log-normal field with standard deviations $\sigma=0.2$, respectively. We observe that \dastar{} and \dT{} both contain a significant fraction of pixels reaching values larger than 1. \textit{Middle column}: Standard deviations of the fields  \dastar{} and \dT{} smoothed over various scales, in the upper and lower panels, respectively. Noticeably, these quantities remain largely below one. \textit{Rightmost column}: the solid lines represent the  power spectra of \U{} and \V, in the upper and lower panels, respectively. Additionally, the dashed lines highlight the spectra of \Utayl{} and \Vtayl, in the upper and lower panels, respectively. We find that the spectra of the Taylor series significantly deviate from the true spectra. This discrepancy is caused by the heavy-tailed distributions of   \dastar{} and \dT.}
    \label{fig:distribution fct}
\end{figure*}

\subsubsection{Lyman-\texorpdfstring{$\mathit{\alpha}$}{} coupling and heating \texorpdfstring{$\mathit{(20< z< 10)}$}{}}
Let us now turn our focus to the cosmic dawn, i.e., the heating and \lyal{} coupling epochs which are taking place between redshift $z\sim 20$ and 10 in our scenarios. Comparing the dashed curves in the upper row with the ones in the middle and bottom rows of Fig.~\ref{fig:PS decomposition 3 models}, we note that the inclusion of all 24 higher-order contributions in \dr{} results in a clear improvement of the model prediction, driving down the excess power predicted by linear theory. However, a significant mismatch persists. The 21cm power spectrum is still overestimated by up to a factor of 2 at k=0.13 Mpc$^{-1}$ around both the heating and the \lyal{} peaks across all three models. At the k=0.64 Mpc$^{-1}$, the mismatch is more pronounced, reaching up to a factor 3-5 in the \cutoff{}, \default{} and \boost{} models. 

The next logical option to investigate is the influence of the remaining higher-order contributions to Eq.~(\ref{eq:dTb nl taylor}). We tested this possibility, finding a significantly worse match to the true signal once all higher-order terms are included. The observed mismatch is mainly driven by the extreme peaks produced by the \da$\times$\dT{} cross-field. Our investigation suggests that the remaining difference between $P_{21}$ and $P_{21, model}$ visible in the bottom row of Fig.~\ref{fig:PS decomposition 3 models} is not due to missing higher-order contributions from the \lyal{} and \Tk{} fields. In what follows, we will demonstrate that the error is instead caused by the Taylor expansion of the \U{} and \V{} fields which can produce very wrong results for the rare pixels where the perturbation criterion breaks down.

\subsubsection{Break-down of perturbation theory due to highly non-Gaussian distributions}\label{sec:non gauss}
In this section, we argue that the failure to correctly capture the 21cm power spectrum with the perturbative approach is caused by a diverging Taylor series at rare peaks where the fields exceed unity. To demonstrate this, let us focus on the first-order Taylor series expansion $U_{\rm \alpha,\,taylor}$ and $V_{\rm k,\,taylor}$ written in Eqs.~(\ref{eq:U expansion}-\ref{eq:V expansion}). These expansion terms are only valid for regions where the temperature and \lyal{} coupling fields stay well below certain values (i.e. where Eq.~\ref{eq:validity Taylor} is satisfied). Since the \xal{} and \Tk{} maps are constructed by the overlap of multiple $1/r^{2}$ profiles centered on halos, they can reach high values in the vicinity of halo centers. In these regions, both \Utayl{} and \Vtayl{} predict very strong peaks while the true \U{} and \V{} cannot exceed one by definition. These artificial peaks contaminate the power spectrum up to large scales ($k\sim0.1$ Mpc$^{-1}$) and are at the origin of the differences between the power spectra of \dTb{} and $dT_{b, \, taylor}$. They form even when the standard deviation of the field is well below unity because the fields are non-Gaussian in nature and feature a tail going to high values. 


To support this last assertion, we measure the power spectra of \U, \V, $U_{\rm \alpha,\,taylor}$ and $V_{\rm k,\,taylor}$ from our maps. We compare these quantities to the scale-dependent variance $\sigma^{2}_{X}(k)$ and the cumulative probability distribution $P(\delta_{X}>\delta)$ of the fields \dastar{} and \dT. We define the scale-dependent variance as $\sigma^{2}_{X}(k) = \langle X_{R}^{2}\rangle -\langle X_{R}\rangle ^{2}$, with $X_{R}$ the X field smoothed over top-hat kernels of radius R, and $k=\pi/R$. The cumulative  distribution function of a field $\delta_{X}$ is computed according to:
\begin{equation}\label{eq:distrib fct}
    P(\delta_{X}>\delta) =  \int_{\delta}^{\infty}P(\delta_{X}=\delta)d\delta \\
\end{equation}
with $P(\delta_{X}=\delta)$ representing the probability that $\delta_{X}$ takes the values $\delta$, computed from the simulations boxes. The quantity $P(\delta_{X}>\delta)$ informs us of the fraction of pixels with values exceeding a threshold $\delta$.

We show our results in Fig.~\ref{fig:distribution fct} for the \default{} model only, as our findings are similar among the three models. The color coding is different than in the previous plots. The blue, orange, and green colors correspond to three different redshifts, $z=15,\,11,\,\text{and}\,7$, respectively. The first row focuses on the properties of the \dastar{} field, and the differences between the power spectrum of \U{} and \Utayl, while the second row focuses on the properties of the \dT{} field, and the differences between \V{} and \Vtayl{}. The first two columns show the cumulative distribution functions and standard deviations of \dastar{} and \dT, in the upper and lower panels, respectively. The solid and dashed black lines visible in the leftmost panels represent the distribution functions of a normal and log-normal field with standard deviation $\sigma=0.2$. The third column compares the power spectra of \U{} and \Utayl{}, and of \V{} and \Vtayl{}, in the upper and lower panel, respectively.

As visible in the third column of Fig.~\ref{fig:distribution fct}, there is an offset between the power spectra of \U{} and \Utayl{} and of \V{} and \Vtayl. This mismatch increases towards small scales and is at the origin of the discrepancy between $P_{\rm 21}$ and $P_{\rm 21}^{\rm lin}$ observed in the middle upper panel of Fig.~\ref{fig:PS decomposition 3 models}. Specifically at redshift $z=15$, $\Delta^{2}_{U_{\rm \alpha, \, taylor}}$ exceeds $\Delta^{2}_{U_{\rm \alpha}}$ by a factor of 2 at k=0.13 Mpc$^{-1}$ and 5 at k=0.64 Mpc$^{-1}$. This discrepancy is equivalent to the deviation between $\Delta^{2}_{\rm 21}$ and $\Delta^{2}_{\rm lin}$. The same conclusions hold at $z=11$ around the heating peak where the mismatch between the spectra of \V{} and \Vtayl{} is of a similar magnitude to the offset between $\Delta^{2}_{\rm 21}$ and $\Delta^{2}_{\rm lin}$. 


The leftmost panels of Fig.~\ref{fig:distribution fct} demonstrate that \dastar{} and \dT{} both present a heavy tail distribution, significantly deviating from a Gaussian or a log-normal field, and characterized by a significantly high fraction of pixels reaching high values $\delta>1$. This distribution of high peaks is correlated to the error on the power spectrum induced by the Taylor series of \U{} and \V, but this relation is non-trivial. For instance, at redshift $z=15$, the \dT{} field is close to a log-normal with variance $\sigma=0.2$, as visible in the lower left panel of Fig.~\ref{fig:distribution fct}. At this redshift, the Taylor expansion is a valid approximation and exhibits a Fourier spectrum converged with its fiducial value, as indicated by the overlap of the solid and dashed blue curve in the lower right panel of Fig.~\ref{fig:distribution fct}. At other redshifts, where the high tail of the distribution flattens, we observe a divergence of $P_{\rm V_{\rm k}}$ compared to $P_{\rm V_{\rm k, \, taylor}}$.

We conclude that despite the individual fields having a variance well below unity, they cannot be treated perturbatively due to their strongly non-Gaussian nature. The presence of strong peaks in the \lyal{} and temperature fields leads to an overestimation of power within the perturbative approach. This bias originates from the invalidity of the Taylor series expansion around the peaks of the individual fields.

\section{Conclusions}\label{sec:conclusions}
Radio interferometers such as the Square Kilometre Array (SKA) will observe the 21cm signal from the neutral hydrogen during the cosmic dawn and the Epoch of reionization. Given the high complexity of coupled hydrodynamical radiative transfer computations, many calculations rely on simplifications and approximations to simulate the 21cm global signal and power spectrum. In this paper, we investigate the validity of a number of approximations commonly made in the literature.

Our analysis is conducted with the one-dimensional radiative-transfer code \beorn{} \cite{Schaeffer:2023rsy} which uses an $N$-body simulation as input and models the \lyal{} coupling, X-ray heating, and the reionization process on a grid painting luminosity profiles around halo centres. The code provides cosmological boxes of the ionized hydrogen fraction (\xHII{}), the kinetic temperature of the gas (\Tk{}), the \lyal{} coupling coefficient (\xal), and the total differential brightness temperature (\dTb{}). We use these simulation boxes to mimic the predictions of various approximations and compare them to the true signal. In order to quantify the sensitivity of our results with respect to the astrophysical modeling, we investigate the three source models \cutoff{}, \default{}, and \boost{} which differ in their stellar-to-halo mass and UV escape fractions for small galaxies (see Fig.~\ref{fig:properties 3models} and corresponding text).

As a first step, we test the assumption of ignoring reionization bubbles when investigating the cosmic dawn, i.e. the epoch of \lyal{} coupling and heating of the neutral gas. We show that disregarding reionization during cosmic dawn has an impact on the 21cm power spectrum up to very high redshifts. Even at early periods when the mean ionization fraction is well below the percent level, order $50\%$ deviations from the true power spectrum can occur in the relevant regime of $k\sim 0.1-1$ Mpc$^{-1}$.

Next, we investigate the very common assumption of a saturated spin temperature during reionization. This corresponds to ignoring all effects from the heating and the \lyal{} coupling period during the reionization process. We find that for all three astrophysical source models considered here, effects from reionization cannot be safely decoupled from the cosmic dawn as the ratio of the mean spin temperature and CMB temperatures ($\bar{T}_{\rm S}/T_{\rm cmb}$) always stays below 100. Even in the late regime where the mean temperature ratio may go to $100>\bar{T}_{\rm S}/T_{\rm cmb}>10$, the assumption of a saturated spin temperature still leads to errors on the 21cm power spectrum of 10-20 \%. This error grows substantially to about an order of magnitude when moving to the regime of $10>\bar{T}_{\rm S}/T_{\rm cmb}>1$. We conclude that the conditions for true spin saturation are hardly ever satisfied for realistic astrophysical source models. 

As a next step, we examine the common approximation of calculating the global differential brightness temperature  \dTbbar{} from the mean of the individual fields that compose the signal. We compare this calculation to the true global signal measured by averaging the \dTb{}-signal from the simulation box. Our analysis reveals a bias of approximately $10\%$ around the dip of the absorption trough in all three astrophysical source models. This error can be reduced when the correlations between the fields are accounted for. We propose a method to include the covariances from the individual clustering terms so that the true signal can be recovered to good precision.  

Finally, we analyze the accuracy of the perturbative approach for computing the 21cm power spectrum. This method is based on a decomposition of the total \dTb{} fluctuations into a series of individual perturbations from the \lyal{}, temperature, gas density, and ionization components. This decomposition is obtained by replacing certain components of the \dTb{} field with their Taylor series truncated to first order, and by neglecting high-order products of individual perturbations. For the epoch of reionization (EoR) we demonstrate (i) the necessity of including higher-order terms in the matter and ionization fields next to the standard linear terms, and (ii) the non-negligible contribution of higher-order terms in the temperature field for models where the epoch of heating overlaps with reionization. A similar study was performed in Refs.~\citep{Lidz:2006vj,Georgiev:2021yvq}, however, without including the impact of temperature and \lyal{} perturbations.

Regarding the epoch of cosmic dawn, the linear perturbation approach overestimates the 21cm power spectrum during \lyal{} coupling and heating by up to an order of magnitude. Including higher-order terms improves the situation somewhat, but differences of a factor of 2-3 remain. We attribute this remaining error to the highly non-Gaussian (heavy-tailed) distribution of the \lyal{} and \Tk{} fields. Despite their variances being well below 1, these fields contain high peaks around large sources where the Taylor series expansion becomes very inaccurate. 

We conclude that analytical approaches based on a perturbative series expansion of the $dT_{\rm b}$ field remain very approximate, a fact that is mainly driven by the nonlinear dependence of the total 21cm signal to the \lyal{} coupling and temperature fields (see Eq.~\ref{eq:dTb}).

\textcolor{black}{The various approximations tested in this paper — separating the epoch of Lyman-$\alpha$ coupling and heating from the epoch of reionization, and using the perturbative approach to calculate the 21cm power spectrum — result in prediction errors significantly larger than the anticipated noise from HERA or SKA observations \citep{Park:2018ljd, Munoz2022, Giri:2022nxq}. These approximations fail to meet the target modelling error of $3\%$ required for SKA-low cosmological constraints to be competitive with Planck, as demonstrated by Ref.~\citep{HMreio}. This underscores the necessity of grid-based simulations that simultaneously model Lyman-$\alpha$ coupling, heating, and reionization to obtain accurate predictions of the total 21cm global signal and, more so, the power spectrum.}


\begin{acknowledgments}
We thank the anonymous referee for their valuable comments and feedback. This work is supported by the Swiss National Science Foundation (SNF) via the grant PCEFP2\_181157.
Nordita is supported in part by NordForsk.
We acknowledge the allocation of computing resources provided by the Swiss National Supercomputing Centre (CSCS) and National Academic Infrastructure for Supercomputing in Sweden (NAISS).
\end{acknowledgments}



\bibliography{21cm_expansion}

\end{document}